\documentclass[prd,
showpacs,showkeys,superscriptaddress,nofootinbib]{revtex4}%

\usepackage{graphicx}
\usepackage{epstopdf}           
\graphicspath{./}

\usepackage{amssymb,amsmath,amsfonts,amsthm}
\usepackage{array,multirow}

\usepackage[colorlinks=true,linkcolor=blue,citecolor=blue]{hyperref}
\usepackage{url}  
\newcommand{\Fig}[1]{Fig.~\ref{#1}}
\newcommand{\Tab}[1]{Table~\ref{#1}}
\newcommand{\Sec}[1]{Sec.~\ref{#1}}

\newcommand{\Eq}[1]{Eq.\,\eqref{#1}}

\newcommand{\Tr}{\operatorname{Tr}}

\renewcommand{\Re}{\operatorname{\mathfrak{Re}}}

\begin{document}


\title{Spectral sum for the color-Coulomb potential \\
in $\boldsymbol{SU(3)}$ Coulomb gauge lattice Yang-Mills theory
}


\author{Y.~Nakagawa}
\affiliation{Research Institute for Information Science and Education,
Hiroshima University, Higashi-Hiroshima, Hiroshima, 739-8521, Japan}

\author{A.~Nakamura}
\affiliation{Research Institute for Information Science and Education,
Hiroshima University, Higashi-Hiroshima, Hiroshima, 739-8521, Japan}

\author{T.~Saito}
\affiliation{Integrated Information Center, 
Kochi University, Kochi, 780-8520, Japan}

\author{H.~Toki}
\affiliation{Research Center for Nuclear Physics, 
Osaka University, Ibaraki, Osaka, 567-0047, Japan}


\begin{abstract}
We discuss the essential role of the low-lying eigenmodes
of the Faddeev-Popov (FP) ghost operator
on the confining color-Coulomb potential
using $SU(3)$ quenched lattice simulations in the Coulomb gauge.
The color-Coulomb potential is expressed as
a spectral sum of the FP ghost operator
and has been explored by partially summing the FP eigenmodes.
We take into account the Gribov copy effects that have a great impact
on the FP eigenvalues and the color-Coulomb potential.
We observe that the lowest eigenvalue vanishes in the thermodynamic limit
much faster than that in the Landau gauge.
The color-Coulomb potential at large distances is governed by
the near-zero FP eigenmodes;
in particular, the lowest one accounts for a substantial portion of
the color-Coulomb string tension comparable to the Wilson string tension.
\end{abstract}



\maketitle

\section{Introduction}

The use of the Coulomb gauge has great advantages in studying the nonperturbative
aspects of Yang-Mills theory, such as color confinement
\cite{GreensiteJ:PPNP51:2003,Zwanziger:1998yf,CucchieriA:AIPCP892:2007}.
In the Coulomb gauge the three dimensionally transverse components and
the time-time component of the gluon propagator do not mix with each other.
Although this makes the perturbative calculation cumbersome,
the confinement phenomenon can be tackled in a comprehensive way.
In QCD gluons play a dual role: confining gluons and confined gluons.
Confining gluons mean that the gluons play the role to confine quarks,
and they cause a strong long-range correlation between quarks distant apart.
Confined gluons indicate that gluons themselves are confined in hadrons,
that is, the gauge fields cannot have a correlation beyond the hadronic scale.
Such complementary aspects of the gluons
can coexist in the Coulomb gauge:
The transverse gluon propagator is suppressed in the infrared region,
which indicates the confinement of gluons, and
the time-time component of the gluon propagator diverges
much stronger in the infrared limit than the free field propagator
\cite{Cucchieri:2000gu, LangfeldK:PRD70:2004, NakagawaY:PRD75:2007, QuandtM:PSLAT2007:2007, Voigt:2007wd,NakagawaY:PRD79:2009,
BurgioG:PRL102:2009,Nakagawa:2009is}.
This interpretation of color confinement differs from the confinement
mechanism in the covariant gauge, in which the highly random fluctuations
of the gauge fields lead the incoherent interference of different paths
of the transport of color charges
\cite{PhysRevD.69.016002}.

The Coulomb gauge Hamiltonian contains the instantaneous interaction
which plays a significant role in the Gribov-Zwanziger scenario
\cite{Zwanziger:1998yf}.
The instantaneous interaction energy between color charges is called
the color-Coulomb potential.
Lattice QCD simulations have showed that
the color-Coulomb potential rises linearly at large distances
and it has $2 \sim 3$ times larger string tension
than the static Wilson potential
\cite{GreensiteJ:PRD67:2003,Greensite:2004ke,NakamuraA:PTP115:2006}.
This is consistent with the Zwanziger's inequality which states that
the instantaneous color-Coulomb potential provides an upper bound
for the static potential
\cite{Zwanziger:2003sh}.
In other words, the necessary condition for the static Wilson potential
being a confining potential is that the color-Coulomb potential is
also a confining potential.
Furthermore, it has been shown that the scaling violation of
the color-Coulomb string tension
is weaker than that of the Wilson string tension
\cite{Greensite:2004ke,Nakagawa:2006fk}.

The confining nature of the color-Coulomb potential is attributed to
the accumulation of the low-lying eigenmodes of
the Faddeev-Popov (FP) ghost operator.
As pointed out by Gribov, the linear gauges such as Coulomb gauge
or Landau gauge do not fix the gauge completely
\cite{Gribov:1978wm,Singer:1978dk}.
There remain gauge-equivalent configurations
after imposing the gauge fixing condition on the gauge fields.
In order to select only one representative from each gauge orbit
we have to restrict the gauge configurations to the so-called Gribov region
(or more precisely, the fundamental modular region)
where the Faddeev-Popov operator is positive.
Because of the entropy considerations
\cite{Zwanziger:1993dh},
a typical gauge configuration lies near the Gribov horizon
where the lowest eigenvalue of the FP operator vanishes.
Consequently, the ghost propagator and the color-Coulomb potential
which contains the inverse of the FP operator twice become infrared singular.
Recent lattice QCD simulations have revealed that
the eigenmodes of the FP operator accumulate in the low-lying level
and its density increases with increasing the lattice volume
\cite{Greensite:2005ur,NakagawaY:PRD75:2007}.

In this study, we investigate the impact of the low-lying eigenmodes
of the FP operator on the confining behavior of the color-Coulomb potential.
The Gribov copy effects are taken into account
by generating the gauge copies and by selecting the representatives
that give the smallest value of the minimizing functional.
We exploit the spectral representation of the FP ghost Green's function.
Because the color-Coulomb potential can be expressed in terms of
the Green's function of the FP operator,
the color-Coulomb potential is explicitly written as a spectral sum
of the FP ghost eigenmodes.
We examine how the long-distance behavior of the color-Coulomb potential
is governed by the low-lying FP eigenmodes.

The organization of the paper is as follows.
In the next section we express the color-Coulomb potential
as the spectral sum of FP ghost eigenmodes.
Section \ref{sec:Results} is devoted to explain the technical details of the simulations
and to show the results of our numerical simulations.
The effects of the Gribov copies on the FP eigenvalues are examined
in \Sec{sec:CopiesOnEigenvalue}.
In \Sec{sec:LowLyingEVs}, the lowest FP eigenvalue will be shown to
vanish in the infinite volume limit faster than that in the Landau gauge.
The color-Coulomb potential obtained by the partial spectral sum and
the color-Coulomb string tension are given in
Secs. \ref{sec:colorCoulomb} and \ref{sec:StringTension}.
We will see that the low-lying modes account for the large portion
of the color-Coulomb string tension,
and the lowest eigenmode has a substantial contribution to the spectral sum.
The results for the weight factor which controls the contribution of
each eigenmode to the spectral sum are given in \Sec{sec:WeightFactor}.
The correlation function of the FP eigenmodes
which is responsible for the distance dependence
of each eigenmode is examined in \Sec{sec:Correlation}.
Finally, we present conclusions in \Sec{sec:Summary}.

\section{Spectral sum for the color-Coulomb potential}
\label{sec:SpectralSum}

In the Coulomb gauge, the Hamiltonian can be decomposed into
the transverse part and the instantaneous part
\cite{Cucchieri:2000hv};
\begin{eqnarray}
H &=& \frac{1}{2}\int d^3x\left\{(E_i^{\textrm{tr}})^2+B_i^2\right\} \nonumber \\
&+& \frac{1}{2}\int d^3y\int d^3z\rho^a(\vec{y}, t)
\mathcal{V}^{ab}(\vec{y},\vec{z}; A^{\textrm{tr}})\rho^b(\vec{z}, t).
\end{eqnarray}
$E_i$ are the transverse components of the color-electric field and
$B_i^a \equiv \epsilon_{ijk}F_{jk}/2$ the color-magnetic fields.
$\rho$ is the color charge density
\begin{equation}
\rho^a=
gf^{abc}A_i^{b,\textrm{tr}}E_i^{c,\textrm{tr}}+\rho_{\textrm{quark}}^a.
\end{equation}
Color charges interact with each other instantaneously
through the kernel $\mathcal{V}$ which is expressed
in terms of the Green's function $M^{-1}$ of the FP ghost operator
$M^{ab}=-\partial_iD_i^{ab}=
-\delta^{ab}\partial_i^2-gf^{abc}A_i^{c, \textrm{tr}}\partial_i$,
\begin{equation}
\mathcal{V}^{ab}(\vec{y},\vec{z}; A^{\textrm{tr}})
=(M^{-1}[A](-\partial_i^2)M^{-1}[A])^{ab}_{\vec{y},\vec{z}}.
\end{equation}
In the Abelian theory the FP operator is the negative Laplacian
since the structure constants $f^{abc}$ are zero.
Accordingly the instantaneous interaction is reduced
to the well-known Coulomb potential.

As can be seen from the Coulomb gauge Hamiltonian,
the instantaneous interaction between quarks is proportional to
the product of the color charges as the one-gluon exchange;
namely, the instantaneous interaction is purely a two-body interaction
and is proportional to the quadratic color factor
\cite{NakagawaY:PRD77:2008}.
The vacuum expectation value of the instantaneous interaction energy between
a quark and an antiquark located at $\vec{x}$ and $\vec{y}$ is given by
\begin{equation}
V_c(\vec{x}-\vec{y}) = g^2 \vec{T}^a_q \cdot \vec{T}^b_{\bar{q}}
\langle \mathcal{V}^{ab}(\vec{x},\vec{y}; A^{\textrm{tr}}) \rangle,
\end{equation}
and we call $V_c$ the color-Coulomb potential.
Here $\langle \cdot \rangle$ denotes the Monte Carlo average and
$T_{q(\bar{q})}^a$ are the generators of the color-$SU(3)$ group.
In the color-singlet channel the color-Coulomb potential is
\begin{equation}
V_c^{\textrm{singlet}}(\vec{x}-\vec{y}) = g^2 \frac{-C_f}{N_c^2-1}
\langle \mathcal{V}^{aa}(\vec{x},\vec{y}; A^{\textrm{tr}}) \rangle,
\end{equation}
where $C_f=4/3$ is the Casimir invariant in the fundamental representation
of the color-$SU(3)$ group and $-C_f$ is the color factor in the singlet channel.
$N_c$ is the number of colors.
In this paper we are interested in only the color-singlet channel and for simplicity
we denote the color-Coulomb potential in the singlet channel by $V_c$.

Our definition of the color-Coulomb potential is slightly different from
what appeared in
\cite{Cucchieri:2000hv}
where the color-Coulomb potential is given
as the instantaneous part of the time-time component of the gluon  propagator,
\begin{eqnarray}
D_{44}(x-y)
&=& \textrm{Tr}\langle A_4(\vec{x},x_4)A_4(\vec{y},y_4) \rangle \nonumber \\
&=& V'_c(|\vec{x}-\vec{y}|)\delta(x_4-y_4) + P(x-y).
\end{eqnarray}
$V'_c$ can be written as
\begin{equation}
V'_c(|\vec{x}-\vec{y}|)\delta^{ab}
= g^2
\langle \mathcal{V}^{ab}(\vec{x},\vec{y}; A^{\textrm{tr}}) \rangle.
\end{equation}
Therefore, our definition of the color-Coulomb potential differs from that in
\cite{Cucchieri:2000hv}
by the color factor $-C_f$.

The Green's function of the FP operator can be expanded explicitly in terms of
the eigenfunctions $\phi_n$ and the eigenvalues $\lambda_n$ of the FP operator as
\begin{equation}\label{GhostModeSum}
(M^{-1}[A])^{ab}(\vec{x},\vec{y})
= \sum_n \frac{\phi_n^{\ast a}(\vec{x})\phi_n^b(\vec{y})}{\lambda_n},
\end{equation}
where the sum extends over the whole eigenmodes besides the trivial zero modes.

In the same way, the color-Coulomb potential can be written as
a spectral sum of the ghost eigenmodes,
\begin{widetext}
\begin{equation}
V_c(\vec{x}-\vec{y})
= g^2\frac{-C_f}{N_c^2-1}
\left\langle \sum_{n,m}\phi_n^{\ast a}(\vec{x})\phi_m^a(\vec{y})
\frac{\int d^3z \phi^c_n(\vec{z})(-\partial_i^2)\phi^{\ast c}_m(\vec{z})}
{\lambda_n\lambda_m}
\right\rangle.
\end{equation}
\end{widetext}
We notice that the distance dependence of the color-Coulomb potential
comes from the product of the eigenmodes
$\phi_n^{\ast a}(\vec{x})\phi_m^b(\vec{y})$.
The remaining part,
\begin{equation}
\omega_{nm}
=\frac{\int d^3z \phi^c_n(\vec{z})(-\partial_i^2)
\phi^{\ast c}_m(\vec{z})}{\lambda_n\lambda_m},
\end{equation}
gives a weight of the contributions from each eigenmode.
In the Abelian theory, the eigenvalue is $1/\vec{p}\,{}^2$ and
the eigenfunctions are the plane waves
since the FP operator is the negative Laplacian.
Therefore the weight factor becomes diagonal
and is proportional to the inverse of the FP eigenvalue,
$\omega_{nm} \propto \delta_{nm}/\lambda_n$.

In the Gribov-Zwanziger scenario
the FP eigenmodes get accumulated in the low-lying modes
\cite{Zwanziger:1993dh},
and as the volume increases
the lowest eigenvalue of the FP operator in the Yang-Mills theory
goes toward zero faster than that in the Abelian theory
where the lowest eigenvalue of the FP operator
on a lattice is $4\sin^2(\pi/L)$, see \Sec{sec:LowLyingEVs}.
Accordingly we expect that the lowest component of
the weight factor $\omega_{11}$ becomes significantly large
in the Yang-Mills theory and
the color-Coulomb potential is dominated by the lowest FP eigenmode.

\section{Numerical simulations}
\label{sec:Results}

The lattice configurations $U=\{U_{\mu}(x)\}$ are generated
by the Cabibbo-Marinari pseudo-heat-bath algorithm \cite{CabibboN:PLB119:1982}
with the Wilson plaquette action on several lattice volumes and lattice couplings.
Ten thousand sweeps are discarded for thermalization and
a measurement has been done every 100 sweeps.
The lattice parameters used in this study are summarized in \Tab{tab:setup}.

The Coulomb gauge condition
\begin{equation}
\partial_iA_i(\vec{x},t) = 0
\end{equation}
does not mix the spatial components and the temporal component
of the gauge fields, and each time slice is gauge-fixed independently.
In this work, we adopt the Wilson-Mandula iterative method
with the Fourier acceleration to fix to the Coulomb gauge
\cite{Davies:1987vs}; namely,
we minimize the functional
\begin{equation}\label{minimizeF}
F_U[g] = \frac{1}{L^3} \sum_{\vec{x}} \Re\Tr
\left( 1 - \frac{1}{3}g^{\dagger}(\vec{x},t)
U_{\mu}(\vec{x},t)g(\vec{x}+\vec{i},t) \right)
\end{equation}
with respect to the gauge transformation $g(\vec{x},t) \in SU(3)$
and find the local minimum of $F_U[g]$.
Here $L^3$ is the spatial lattice volume.
The gauge fields $A_{\mu}(\vec{x},t)$ are defined as
\begin{equation}
A_{\mu}(\vec{x},t) = \left.
\frac{U_{\mu}(\vec{x},t)-U^{\dagger}_{\mu}(\vec{x},t)}{2iag_0}
\right|_{\textrm{traceless}},
\end{equation}
and the gauge fields at the minima of $F_U[g]$ satisfy
the transversality condition $\nabla_i A_i(\vec{x},t)=0$ where
$\nabla$ is the lattice backward difference.
We stop the iterative gauge fixing if the violation of the transversality
becomes less than $10^{-14}$;
\begin{equation}
\theta = \frac{1}{(N_c^2-1)L^3}\sum_{a,\vec{x}}
(\nabla_i A_i^a(\vec{x},t))^2 < 10^{-14}.
\end{equation}
This stopping criterion is applied for each time slice.
We refer the configurations gauge-fixed in this way to \textit{first copies}.

In order to examine the effects of Gribov copies,
we generate 30 gauge copies by performing
the random gauge transformation for each time slice.
Then, we find the minimum of the functional (\ref{minimizeF}).
Configurations giving a lowest value of the minimizing functional
among the gauge copies for all time slices are called \textit{best copies},
candidates of gauge configurations in the fundamental modular region.
In this way, we can prepare the configurations that are close to
the fundamental modular region and discuss the Gribov copy effects.

The effects of the Gribov copies on the color-Coulomb potential
obtained by inverting the FP operator
in the conjugate gradient method have been discussed in
\cite{VoigtA:PRD78:2008}.
It has been clarified that the Gribov copies tremendously affect
the color-Coulomb potential at small momenta.
In the following, we shall see that the color-Coulomb string tension
extracted from the spectral sum of the FP eigenmodes is also influenced
by the Gribov copies.

\begin{table}[h]
\caption{
The lattice couplings, the lattice volumes, the number
of configurations used to evaluate the FP eigenvalues
and the corresponding eigenfunctions.
$N_{cp}$ refers to the number of random gauge copies
generated for each time slice to investigate the Gribov copy effects.
}
\begin{center}
\begin{tabular}{ccccccc}
\hline
$\beta$ & $L^4$ & $a^{-1}$ [GeV] &
$a$ [fm] & $V$[fm$^4$] & \# of confs. & $N_{cp}$ \\
\hline \hline
\multirow{1}{*}{5.70}
&  8$^4$ & 1.160 & 0.1702 & 1.36$^4$ & 100 & 30 \\
& 16$^4$ & 1.160 & 0.1702 & 2.72$^4$ & 100 & 30 \\
\hline
\multirow{5}{*}{5.80}
&  8$^4$ & 1.446 & 0.1364 & 1.09$^4$ & 100 & 30 \\
& 12$^4$ & 1.446 & 0.1364 & 1.64$^4$ & 100 & 30 \\
& 16$^4$ & 1.446 & 0.1364 & 2.18$^4$ & 100 & 30 \\
& 20$^4$ & 1.446 & 0.1364 & 2.73$^4$ & 100 & 30 \\
& 24$^4$ & 1.446 & 0.1364 & 3.27$^4$ & 100 & 30 \\
\hline
\multirow{2}{*}{6.00}
&  8$^4$ & 2.118 & 0.0931 & 0.74$^4$ & 100 & 30 \\
& 16$^4$ & 2.118 & 0.0931 & 1.49$^4$ & 100 & 30 \\
& 24$^4$ & 2.118 & 0.0931 & 2.23$^4$ & 100 & 30 \\
\hline
\multirow{1}{*}{6.20}
&   8$^4$ & 2.914 & 0.0677 & 0.54$^4$ & 100 & 30 \\
&  16$^4$ & 2.914 & 0.0677 & 1.08$^4$ & 100 & 30 \\
&  24$^4$ & 2.914 & 0.0677 & 1.63$^4$ & 100 & 30 \\
\hline
\end{tabular}
\label{tab:setup}
\end{center}
\end{table}

In the Coulomb gauge the FP operator is a $8L^3 \times 8L^3$ matrix
on a lattice and a purely spatial quantity.
It can be expressed in terms of the spatial link variables $U_i(x)$ as
\begin{widetext}
\begin{equation}\label{FPmat}
M_{\vec{x}\vec{y}}^{ab} =
\sum_{i}\mathfrak{Re}\textrm{Tr}
\left[\{T^a, T^b\}\left(U_{i}(\vec{x},t)+U_{i}(\vec{x}-\hat{i},t)\right)
\delta_{\vec{x},\vec{y}}
- 2T^bT^a U_{i}(\vec{x},t) \delta_{\vec{y},\vec{x}+\hat{i}}
- 2T^aT^bU_{i}(\vec{x}-\hat{i})\delta_{\vec{y},\vec{x}-\hat{i}} \right].
\end{equation}
\end{widetext}
Since the FP matrix is the Hessian matrix
associated with the functional (\ref{minimizeF}),
the eigenvalues we obtain are positive.
In $SU(3)$ Yang-Mills theory there are eight trivial zero modes 
associated with the spatially constant eigenfunctions.
We exclude these modes from the spectral sum of the color-Coulomb potential.

We solve the eigenvalue equation
\begin{equation}
M\phi_n = \lambda_n\phi_n,
\end{equation}
and evaluate low-lying 300 eigenvalues and the corresponding eigenvectors
including the trivial zero modes using the ARPACK package
\cite{ARPACK}.
The spectral representation of the color-Coulomb potential
in the singlet channel is
\begin{equation}
V_c(R) = g^2 \frac{-C_f}{N_c^2-1} \left\langle
\sum_{k,l=n_{\textrm{min}}}^{n_\textrm{max}}
\phi_k^{\ast a}(\vec{x})\phi_l^a(\vec{y})\omega_{kl}
\right\rangle,
\end{equation}
where $R=|\vec{x}-\vec{y}|$.
We investigate the behavior of the color-Coulomb potential
by changing the lower and the upper limits of the spectral sum,
$n_{\textrm{min}}$ and $n_{\textrm{max}}$.
Statistical errors are estimated by the jackknife method.

\subsection{Effects of the Gribov copies on the FP eigenvalues}
\label{sec:CopiesOnEigenvalue}

\begin{figure*}[htbp]
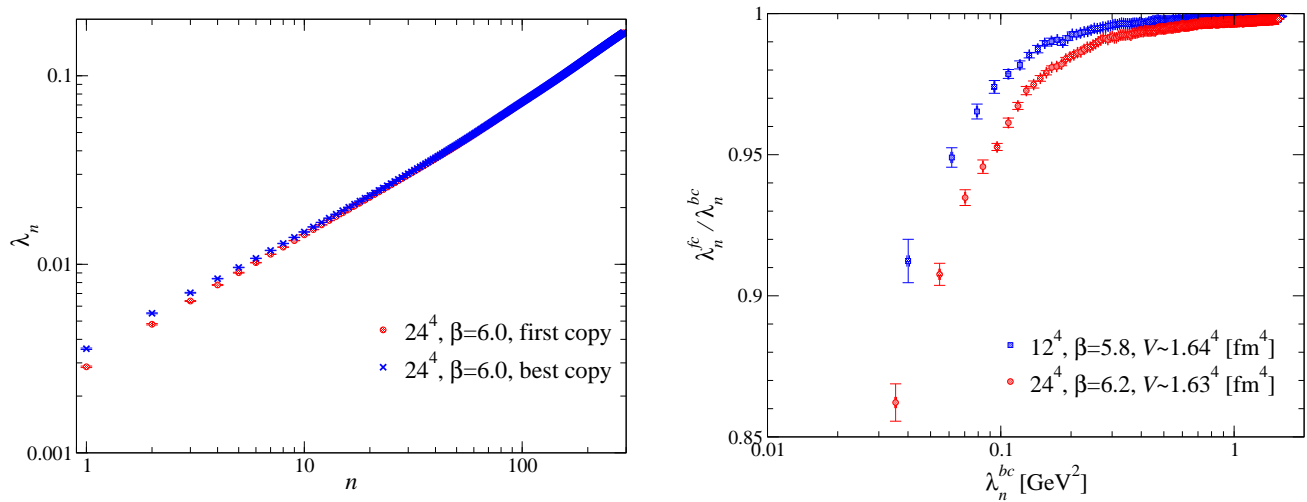

\begin{minipage}{0.46\hsize}\begin{center}
\resizebox{1.\textwidth}{!}
{\includegraphics{fig1a}}
\end{center}\end{minipage}
\hspace{0.03\hsize}
\begin{minipage}{0.46\hsize}\begin{center}
\resizebox{1.\textwidth}{!}
{\includegraphics{fig1b}}
\end{center}\end{minipage}
\hspace{0.03\hsize}
\caption{
(Left) The eigenvalues of the FP operator on the $24^4$ lattice
at $\beta=6.0$ for the first copies and the best copies.
(Right) The ratio $\lambda_n^{fc}/\lambda_n^{bc}$ of the FP eigenvalues
on almost the same physical volume is plotted
as a function of $\lambda_n^{bc}$.
}
\label{GribovCopy_EigenValue}
\end{figure*}

We discuss the effects of the Gribov copies on the eigenvalues
of the FP operator.
In the left panel of \Fig{GribovCopy_EigenValue},
the lowest nontrivial 292 eigenvalues evaluated on the first copies and
the best copies on the $24^4$ lattice at $\beta=6.0$ are plotted.
We observe that the first copies underestimate the FP eigenvalues and
the discrepancies are large at small eigenvalues.
The ratio $\lambda_n^{fc}/\lambda_n ^{bc}$ of the FP eigenvalues
on almost the same physical volume is plotted in the right panel
of \Fig{GribovCopy_EigenValue}.
Here $\lambda_n^{fc}$ refers to the FP eigenvalue for the first copies
and $\lambda_n^{bc}$ for the best copies.
We see that the ratio decreases as we get close to the critical limit
(i.e., the thermodynamic limit and the continuum limit).
The effects of the Gribov copies cause more than 10 $\%$ systematic errors
for the lowest eigenvalue on the $24^4$ lattice.
This indicates that we have to pay adequate attention to
the Gribov copy effects on the FP eigenvalues.

The reason why the best copies give larger eigenvalues than
the first copies can be explained as follows.
Except for the trivial zero modes,
the FP eigenvalues are strictly positive for the trivial vacuum $A=0$.
As we come close to the first Gribov horizon,
the lowest eigenvalue decreases and becomes zero
for configurations on the first Gribov horizon.
The fundamental modular region is a subset
of the Gribov region and its boundary does not necessarily
coincide with the Gribov horizon,
although it has a common boundary with the Gribov region
\cite{Zwanziger:1993dh}.
Therefore, the fundamental modular region has a small overlap
with the Gribov horizon compared to the Gribov region.
It implies that the FP eigenvalues for the best copies are larger than
that for the first copies, which is seen in our result.

\subsection{Lowest eigenvalues of the FP operator}
\label{sec:LowLyingEVs}

\begin{figure}[htbp]\begin{center}
\resizebox{0.46\textwidth}{!}
{\includegraphics{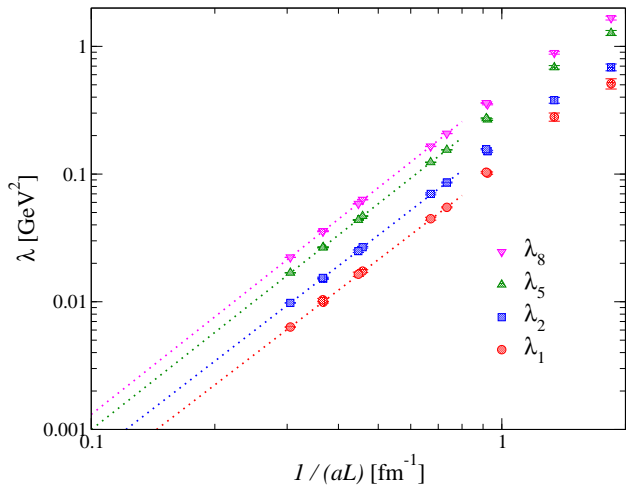}}
\caption{
The lowest eigenvalues, $\lambda_i$ ($i=1, 2, 5, 8$),
for the best copies are plotted as a function of
the inverse of the spatial lattice extent.
The dotted lines correspond to the fitted results.
}
\label{fig:LowestEVs}
\end{center}\end{figure}

\begin{table}[htbp]
\caption{
The fitted results for the lowest eigenvalues of the FP operator.
$(1/aL)_{\textrm{max}}$ [fm$^{-1}$] is the maximum value
of the fitting range.
}
\begin{center}
\begin{tabular}{ccccc}
\hline
$\lambda_i$ & $(1/aL)_{\textrm{max}}$ [fm$^{-1}$]
& $b$ & $c$ & $\chi^2/N_{\textrm{DF}}$ \\
\hline\hline
\multirow{3}{*}{$\lambda_1$}
& 0.6 & 0.121(6) & 2.49(4) & 1.02 \\
& 0.7 & 0.121(4) & 2.48(3) & 0.77 \\
& 0.8 & 0.118(1) & 2.46(1) & 0.76 \\
\hline
\multirow{3}{*}{$\lambda_2$}
& 0.6 & 0.180(6) & 2.46(3) & 0.67 \\
& 0.7 & 0.186(4) & 2.48(2) & 0.86 \\
& 0.8 & 0.185(2) & 2.48(1) & 0.72 \\
\hline
\multirow{3}{*}{$\lambda_5$}
& 0.6 & 0.331(6) & 2.51(2) & 0.30 \\
& 0.7 & 0.339(3) & 2.53(1) & 0.74 \\
& 0.8 & 0.338(2) & 2.53(1) & 0.61 \\
\hline
\multirow{3}{*}{$\lambda_8$}
& 0.6 & 0.461(6) & 2.55(1) & 1.17 \\
& 0.7 & 0.455(3) & 2.54(1) & 1.15 \\
& 0.8 & 0.454(2) & 2.54(1) & 0.93 \\
\hline
\end{tabular}
\end{center}
\label{tab:lowestEVs}
\end{table}

At the zeroth order of the coupling, the FP operator is the negative
Laplacian, $M^{ab} = -\delta^{ab} \partial_i^2$, and
there are $(N_c^2-1)$ nontrivial lowest eigenmodes taking the value
\begin{equation}
\lambda = \left[\frac{2}{a} \sin \left( \frac{\pi}{L} \right) \right]^2,
\end{equation}
which are degenerated.
In the infinite volume limit, this eigenvalue approaches zero as
\begin{equation}\label{AbelianLowest}
\left[\frac{2}{a} \sin \left( \frac{\pi}{L} \right) \right]^2
\longrightarrow \left( \frac{1}{aL} \right)^2.
\end{equation}
We investigate the volume dependence of
the lowest eigenvalues of the FP operator and see if
they vanish faster than that for the negative Laplacian,
\Eq{AbelianLowest}.

In \Fig{fig:LowestEVs}, the lowest eigenvalues,
$\lambda_i$ ($i=1, 2, 5, 8$), for the best copies are plotted
as a function of the inverse of the spatial extent of the lattice.
We find that the eigenvalues decrease with increasing the lattice
volumes.
Moreover, the data points seem to lie on straight lines in the log-log plot.
In order to explore the volume dependence of the eigenvalues,
we fit the data with the function
\begin{equation}
\lambda_i = b\left(\frac{1}{aL}\right)^c,
\end{equation}
in the range $(1/aL) \le (1/aL)_{\textrm{max}}$.
The fitted parameters are given in Table \ref{tab:lowestEVs}.
The corresponding fitted functions for
$(1/aL)_{\textrm{max}}=0.8$ [fm$^{-1}$]
are drawn in \Fig{fig:LowestEVs} by the dotted lines.
We observe that the fitting works well and
the exponent $c$ is larger than 2 in all the cases.
It implies that not only the lowest FP eigenvalue approaches zero
faster than that of the lattice Laplacian operator, but the low-lying
eigenvalues also do.
Our result is consistent with the hypothesis in the Gribov-Zwanziger scenario
that the measure of the path integral is concentrated on
the part of the horizon where ``all horizons are one horizon"
\cite{Zwanziger:1992qr}.

This has also been observed in the Landau gauge
\cite{SternbeckA:PRD73:2006}.
However, the fitting results have been obtained as
2.16(4), 2.24(5), and 2.45(4) for
$\lambda_1$, $\lambda_2$, and $\lambda_5$, respectively.
It means that the lowest eigenvalue of the FP operator
in the Coulomb gauge vanishes faster than that in the Landau gauge.
Since the ghost propagator can be expanded by the FP eigenmodes
as \Eq{GhostModeSum}, it may be expected to be enhanced
in the Coulomb gauge compared to the Landau gauge.

\subsection{Spectral sum of the color-Coulomb potential}
\label{sec:colorCoulomb}

\begin{figure*}[htbp]
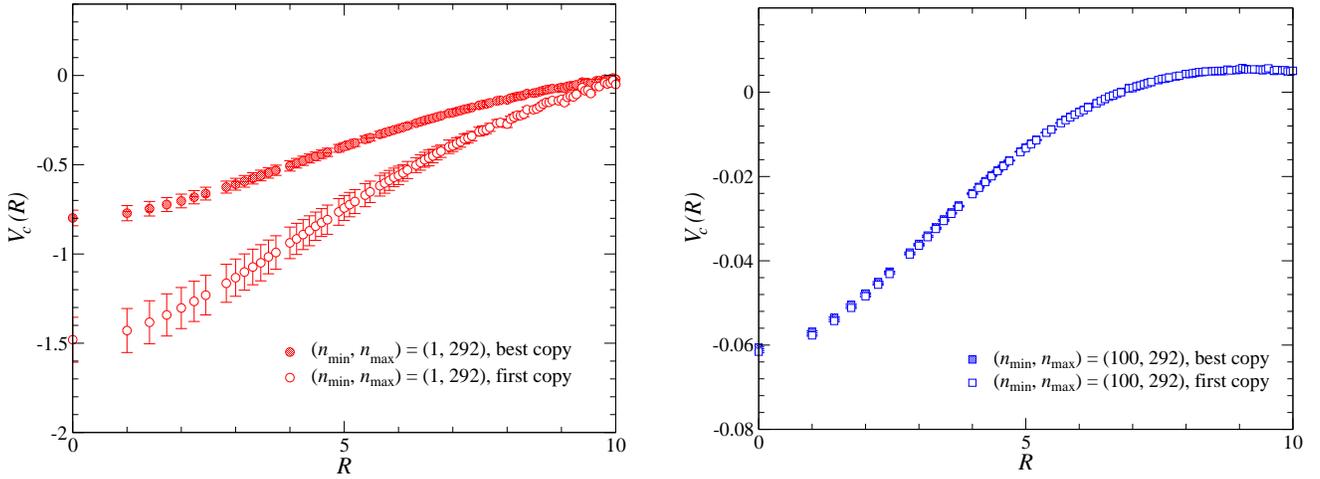

\begin{minipage}{0.46\hsize}\begin{center}
\resizebox{1.\textwidth}{!}
{\includegraphics{fig3a}}
\end{center}\end{minipage}
\hspace{0.03\hsize}
\begin{minipage}{0.46\hsize}\begin{center}
\resizebox{1.\textwidth}{!}
{\includegraphics{fig3b}}
\end{center}\end{minipage}
\hspace{0.03\hsize}
\caption{
The spectral sum of the color-Coulomb potential
for the first and the best copies on the $24^4$ lattice
at $\beta=6.0$ in lattice units.
The lower-upper limit of the sum is
$(1, 292)$ on the left figure and $(100, 292)$ on the right figure.
}
\label{ColorCoulomb_frstbest}
\end{figure*}

We here discuss the spectral sum of
the color-Coulomb potential.
Figure \ref{ColorCoulomb_frstbest} illustrates the effects of
Gribov copies on the color-Coulomb potential.
In the left figure, the result for
$(n_{\textrm{min}},n_{\textrm{max}}) = (1, 292)$ is shown.
We find that the color-Coulomb potential becomes shallow
for the best copies compared to the first copy result.
The discrepancy between the first and the best copies
reaches about 200\% at $R=0$.
In addition to the absolute value,
the statistical errors are also influenced by the Gribov copies;
namely, they are reduced for the best copies.
Accordingly, it is indispensable to take into account the effects
of the Gribov copies to calculate the color-Coulomb potential.

In the right panel of \Fig{ColorCoulomb_frstbest},
the color-Coulomb potential for
$(n_{\textrm{min}},n_{\textrm{max}}) = (100, 292)$ is depicted.
We see that the spectral sum of the color-Coulomb potential is
less affected by the Gribov copies if we exclude the low-lying
modes from the summation.
This result is quite consistent with those
discussed in the previous subsection, that is,
the higher FP eigenvalues are less influenced by the Gribov copies.

\begin{figure*}[htbp]
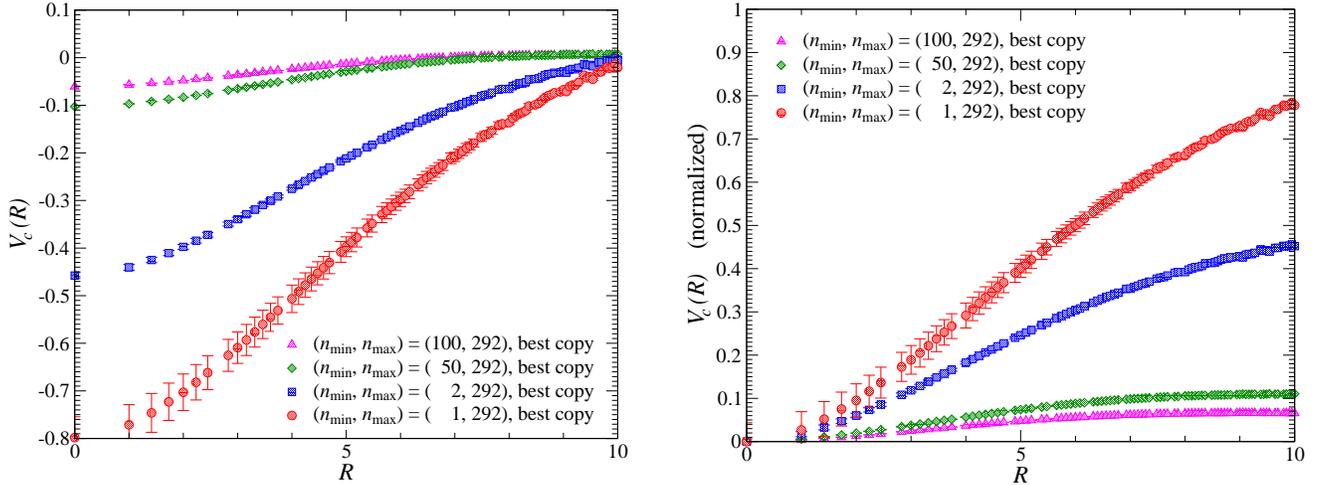

\begin{minipage}{0.46\hsize}\begin{center}
\resizebox{1.\textwidth}{!}
{\includegraphics{fig4a}}
\end{center}\end{minipage}
\hspace{0.03\hsize}
\begin{minipage}{0.46\hsize}\begin{center}
\resizebox{1.\textwidth}{!}
{\includegraphics{fig4b}}
\end{center}\end{minipage}
\hspace{0.03\hsize}
\caption{
(Left) 
The partially summed color-Coulomb potential as a function of $R$
for the best copies in lattice units.
Circles, squares, diamonds, and triangles correspond to
$(n_{\textrm{min}},n_{\textrm{max}}) =
(1,292), (2,292), (50,292), (100,292)$, respectively.
(Right)
The partially summed color-Coulomb potential normalized to be 0 at $R=0$
in order to make the distance dependence more visible.
}
\label{SpSumColorCoulomb}
\end{figure*}

The partial spectral sum of the color-Coulomb potential for the best copies
is plotted in \Fig{SpSumColorCoulomb}.
The right panel illustrates the color-Coulomb potential
normalized to be 0 at $R=0$
in order to make the distance dependence more visible.
By comparing data for $n_{\textrm{min}}=1$ with different $n_{\textrm{min}}$,
we observe that the exclusion of the near-zero modes substantially reduces
the slope and the absolute value of the color-Coulomb potential.
This is because the low-lying components of the weight factor, $\omega_{nm}$,
are quite large and $\omega_{nm}$ rapidly decrease by increasing $n$ or $m$,
as we shall show later.
We should mention that the long-distance behavior of the potential
is governed by only a small fraction of ghost eigenmodes
since the total number of the FP eigenvalues on the $24^4$ lattice is $8L^3=110592$.

Data for $(n_{\textrm{min}}, n_{\textrm{max}})=(50,292)$ and $(100,292)$
in \Fig{SpSumColorCoulomb} suggest that
the inclusion of the eigenmodes in the range $50 \le n \le 99$
changes only the short-distance behavior
of the color-Coulomb potential and the long-range part is not affected.  
From \Fig{SpSumColorCoulomb} we see that the absolute value
of the color-Coulomb potential at large distances decreases
by increasing $n_{\textrm{min}}$ and it gets close to zero.
In other words, the non low-lying modes contribute only to
the short distant part of the potential and its long distant part
is not altered by them.
Thus the inclusion of the non low-lying FP eigenmodes changes only
the short-distance behavior of the color-Coulomb potential.
  
\subsection{Color-Coulomb string tension}
\label{sec:StringTension}

\begin{table*}[htdp]
\caption{
The result of fitting the color-Coulomb potential
with a straight line,
$V(R) = K_c R + c$ in the range $4 \le R \le 7$.
The Wilson string tension $K_W$ is taken from
\cite{BaliGS:PRD47:1993}.
\vspace{2mm}
}
\begin{minipage}{0.49\hsize}
\begin{center}\begin{tabular}{ccccc}
\hline\hline
$n_{\textrm{max}}$ & $K_c^{fc,\,\beta=6.0}$ &
$\chi^2/N_{\textrm{DF}}$ & $K_W^{\beta=6.0}$ & $K_c / K_W$ \\
\hline
  1 & 0.1120( 9) & 0.19 &            & 2.18(12) \\
 16 & 0.1490(12) & 0.16 &            & 2.90(16) \\
 30 & 0.1624(12) & 0.19 &            & 3.17(18) \\
 50 & 0.1668(12) & 0.20 & 0.0513(25) & 3.25(18) \\
100 & 0.1730(12) & 0.23 &            & 3.37(19) \\
200 & 0.1782(12) & 0.29 &            & 3.47(19) \\
292 & 0.1804(12) & 0.36 &            & 3.52(19) \\
\hline\hline
\end{tabular}\end{center}
\begin{center}\begin{tabular}{ccccc}
\hline\hline
$n_{\textrm{max}}$ & $K_c^{bc,\,\beta=5.8}$ &
$\chi^2/N_{\textrm{DF}}$ & $K_W^{\beta=5.8}$ & $K_c / K_W$ \\
\hline
  1 & 0.1050(183) & 0.08 &            & 0.96(19) \\
  8 & 0.1249( 65) & 0.90 &            & 1.15( 8) \\
 30 & 0.1477( 65) & 1.13 &            & 1.36( 8) \\
 50 & 0.1550( 65) & 1.18 & 0.1090(20) & 1.43( 9) \\
100 & 0.1646( 65) & 1.28 &            & 1.51( 9) \\
200 & 0.1720( 65) & 1.65 &            & 1.58( 9) \\
292 & 0.1749( 65) & 2.08 &            & 1.60( 9) \\
\hline\hline
\end{tabular}\end{center}
\end{minipage}
\begin{minipage}{0.49\hsize}
\begin{center}\begin{tabular}{ccccc}
\hline\hline
$n_{\textrm{max}}$ & $K_c^{bc,\,\beta=6.0}$ &
$\chi^2/N_{\textrm{DF}}$ & $K_W^{\beta=6.0}$ & $K_c / K_W$ \\
\hline
  1 & 0.0413(22) & 0.41 &            & 0.81( 8) \\
  8 & 0.0676(38) & 0.26 &            & 1.32(14) \\
 30 & 0.0798(37) & 0.36 &            & 1.56(15) \\
 50 & 0.0841(37) & 0.38 & 0.0513(25) & 1.64(15) \\
100 & 0.0901(37) & 0.46 &            & 1.76(16) \\
200 & 0.0953(37) & 0.78 &            & 1.86(16) \\
292 & 0.0975(37) & 1.15 &            & 1.90(16) \\
\hline\hline
\end{tabular}\end{center}
\begin{center}\begin{tabular}{ccccc}
\hline\hline
$n_{\textrm{max}}$ & $K_c^{bc,\,\beta=6.2}$ &
$\chi^2/N_{\textrm{DF}}$ & $K_W^{\beta=6.2}$ & $K_c / K_W$ \\
\hline
  1 & 0.0276(24) & 0.21 &            & 1.05(12) \\
  8 & 0.0389(14) & 1.92 &            & 1.48( 9) \\
 30 & 0.0469(14) & 2.23 &            & 1.79( 9) \\
 50 & 0.0498(14) & 2.03 & 0.0262(06) & 1.90(10) \\
100 & 0.0544(14) & 1.35 &            & 2.08(10) \\
200 & 0.0587(14) & 1.46 &            & 2.24(10) \\
292 & 0.0609(14) & 2.56 &            & 2.32(11) \\
\hline\hline
\end{tabular}\end{center}
\end{minipage}
\label{Tab:fit_saturation}
\end{table*}

\begin{figure}[htbp]\begin{center}
\resizebox{0.46\textwidth}{!}
{\includegraphics{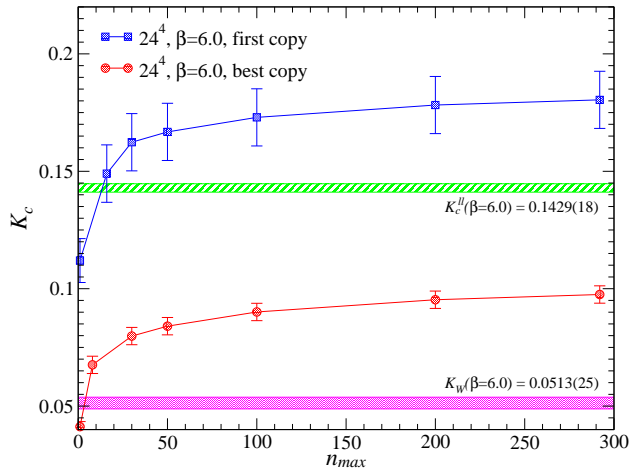}}
\caption{
The color-Coulomb string tension extracted by fitting
the color-Coulomb potential with a straight line
is plotted as a function of $n_{\textrm{max}}$,
the maximum FP mode in the spectral summation.
The squares correspond to the first copy result
on the $24^4$ lattice at $\beta=6.0$, and the circles
represent the best copy one.
The upper band and lower band indicate
the color-Coulomb string tension $K_c$
obtained by measuring the link-link correlator and
the Wilson string tension $K_W$
obtained from the Wilson loop calculation, respectively.
}
\label{SpSum_StrTension_cumulative_fcbc}
\end{center}\end{figure}

In order to reveal how the low-lying FP modes saturate
the color-Coulomb potential at large distances,
we extract the color-Coulomb string tension by fitting the potential
with a straight line and
investigate how the color-Coulomb string tension changes by
varying the maximum mode $n_{\textrm{max}}$ of the spectral sum.
We do not include the Coulomb term, $1/r$, in the fitting function
since the higher FP modes that are responsible for the short distant part
of the color-Coulomb potential are not incorporated in the analysis.
The fitted results are summarized in \Tab{Tab:fit_saturation}.
In this analysis, $n_{\textrm{min}}$ is fixed to be 1.
Thus, the result for $n_{\textrm{max}}=50$ means that
the color-Coulomb potential is reconstructed from the low-lying
50 eigenmodes.

We discuss the effects of the Gribov copies on
the color-Coulomb string tension.
This is illustrated in \Fig{SpSum_StrTension_cumulative_fcbc},
where the first copy and the best copy results are shown.
The string tension $K_c^{ll}$ of the color-Coulomb potential
which is obtained by measuring the link-link correlator
\footnote{
In this section, the color-Coulomb string tension
for the spectral summed potential is referred to as $K_c$
and the string tension of the color-Coulomb potential
obtained from the link-link correlator is referred to as $K_c^{ll}$.
}
and the Wilson string tension $K_W$, the string tension
of the static potential, are also depicted by bands, respectively.
We note that $K_c^{ll}$ is the best copy result and the detail of
the calculation is given in the Appendix.
We find that the first copy result for $K_c$ overestimates $K_c^{ll}$
and it is more than 3 times larger than the Wilson string tension.
By taking into account the Gribov copies,
the color-Coulomb string tension is drastically reduced and it becomes
less than $K_c^{ll}$ although it is still larger than the Wilson string tension.

It may be not a problem that $K_c$ does not agree with $K_c^{ll}$.
The color-Coulomb potential can be calculated in different ways.
One way is to calculate the color-Coulomb potential directly
by inverting the FP matrix or solving the eigenvalue equation
of the FP ghost matrix.
In this method, only the spatial link variables at a fixed time slice
are needed to calculate the color-Coulomb potential.
The other way is to exploit the fact that the color-Coulomb potential is
the instantaneous part of the time-time component of the gluon propagator,
\begin{equation}
D_{44}(\vec{x},t) = V'_c(\vec{x})\delta(t) + P(\vec{x},t),
\end{equation}
and to calculate the correlator of the timelike links at a fixed time slice.
Only the temporal links are needed to calculate the color-Coulomb potential
in this method.
Although these two ways provide a complementary way to obtain
the color-Coulomb potential, the lattice results performed
at finite lattice spacings do not necessarily agree.
Of course, the results should coincide in the continuum limit.

\begin{figure}[htbp]\begin{center}
\resizebox{0.46\textwidth}{!}
{\includegraphics{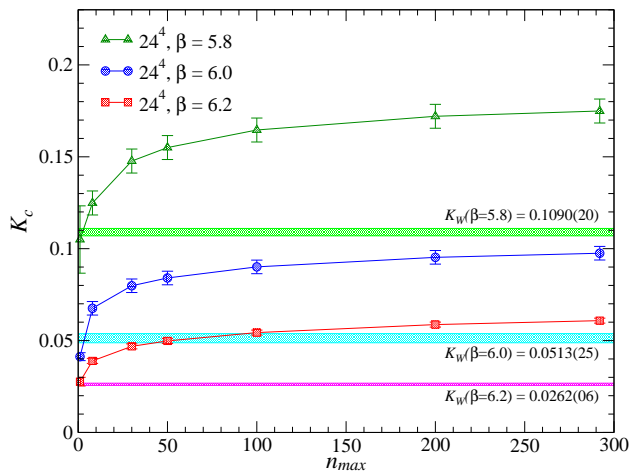}}
\caption{
The color-Coulomb string tension for the best copies
is plotted as a function of $n_{\textrm{max}}$,
the maximum FP mode in the spectral summation.
The Wilson string tension is also shown by the bands.
}
\label{SpSum_StrTension_cumulative}
\end{center}\end{figure}

We next discuss the $n_{\textrm{max}}$ dependence of
the color-Coulomb string tension.
The color-Coulomb string tension for the best copies
at various lattice couplings and
the corresponding Wilson string tension are displayed in
\Fig{SpSum_StrTension_cumulative}.
We see that the color-Coulomb string tension
for the lowest eigenmode is comparable to the Wilson string tension,
indicating that the lowest mode accounts for the large
portion of the string tension.
$K_c$ rapidly increases at small $n_{\textrm{max}}$, and
the rate of increase decreases with $n_{\textrm{max}}$.
The values of $K_c$ for $n_{\textrm{max}}=200$ 
and for $n_{\textrm{max}}=292$ almost agree within
the statistical errors.
Thus, it is expected that the further inclusion of the higher eigenmodes
into the spectral sum does not alter the color-Coulomb string tension
and the long-range behavior of the color-Coulomb potential is
governed by the low-lying eigenmodes.

The ratio of the color-Coulomb string tension to the Wilson string tension
is listed in \Tab{Tab:fit_saturation}.
We observe that the ratio for $n_{\textrm{max}}=292$ increases
with increasing the lattice couplings, ranging from 1.6 to 2.3.
Since the inclusion of the higher eigenmodes,
which we do not take into account,
does not reduce the values of the string tension,
our results exclude the possibility that the color-Coulomb string tension
saturates the Wilson string tension.
Instead, the lowest FP eigenmode gives a large contribution
to the color-Coulomb string tension comparable to the Wilson string tension.

In \cite{LangfeldK:PRD70:2004,VoigtA:PRD78:2008},
the color-Coulomb string tension has been estimated from the zero-momentum
value of the color-Coulomb potential in momentum space.
This method requires simulations on large lattice volumes
to extract a reliable value of the string tension.
However, large lattice simulations of the color-Coulomb potential
are extremely cumbersome since we have to take into account
the Gribov copies, which greatly influence the color-Coulomb potential.
The method we used in this paper is superior in this respect; namely,
the spectral expansion of the color-Coulomb potential in position space
does not require so much large lattices
and only a few hundreds of FP eigenmodes are needed
to estimate the color-Coulomb string tension.

The fact that the color-Coulomb string tension is larger than
the Wilson string tension is physically admissible.
If $K_c$ saturates $K_W$,
the color-Coulomb potential has the same energy
as the ground state energy of the QCD Hamiltonian, that is,
the Wilson static potential.
Since the color-Coulomb potential is the energy of the system
obtained by adding a $q\bar{q}$ pair to the QCD vacuum
\cite{Zwanziger:2003sh},
$K_c = K_W$ would indicate that such a state is also
the eigenstate of the QCD Hamiltonian.
The QCD vacuum does not have a flux tube,
and it also holds for the state,
$|\textrm{QCD vacuum} + q\bar{q} \,\, \textrm{pair} \rangle$.
Hence there is no room for the formation of the flux tube
between quarks if $K_c$ saturates $K_W$.
Moreover, if such a state having no flux tube was the eigenstate
of the QCD Hamiltonian, it could be found by the lattice calculations
as the degenerated ground state of the $q\bar{q}$ system.
Currently, the lattice simulations, however, suggest that
the state with the flux tube is the ground state of the $q\bar{q}$ system
and the state without the flux tube has not been found as the ground
state of the QCD Hamiltonian.
Thus, our result that the color-Coulomb string tension is larger than
the Wilson string tension is quite reasonable.

\subsection{Weight factor $\omega_{nm}$}
\label{sec:WeightFactor}

\begin{figure*}[htbp]
\begin{minipage}{0.46\hsize}\begin{center}
\resizebox{1.2\textwidth}{!}
{\includegraphics{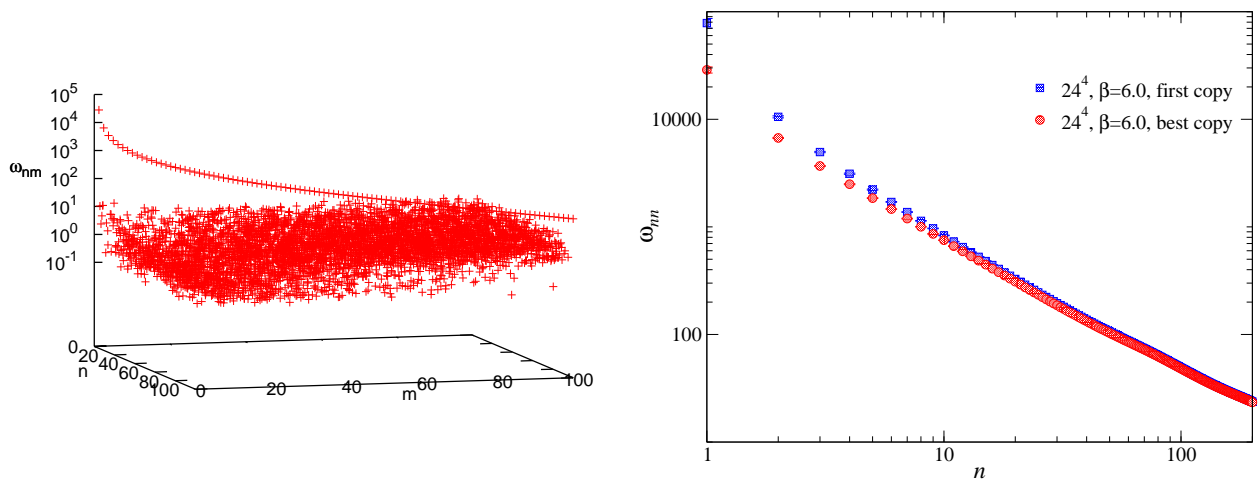}}
\end{center}\end{minipage}
\hspace{0.03\hsize}
\begin{minipage}{0.46\hsize}\begin{center}
\resizebox{1.\textwidth}{!}
{\includegraphics{fig7b}}
\end{center}\end{minipage}
\hspace{0.03\hsize}
\caption{
(Left) $\omega_{nm}$ for the best copies
are plotted as a function of $n$ and $m$
in the range $n,m=1 \sim 100$.
Statistical errors are not shown.
(Right) The diagonal components of the weight factor, $\omega_{nn}$,
are plotted as a function of the label $n$ in the range $1 \le n \le 200$
for both the first copy and the best copy results.
}
\label{WeightFactor}
\end{figure*}

The weight factor $\omega_{nm}$ for the best copies is depicted
in the left panel of \Fig{WeightFactor}.
We observe that the diagonal components are extremely larger
than the off-diagonal components.
This means that the contributions to the spectral sum
of the color-Coulomb potential
from the combinations $\phi_n^{\ast}(\vec{x})\phi_m(\vec{y})$
with different eigenmodes $n \neq m$
are small.
In the Abelian theory the off-diagonal components are exactly zero, 
$\omega_{nm}\sim\delta_{nm}/\lambda_n$.
It is due to the non-Abelian nature of Yang-Mills theory that
the off-diagonal components have finite values,
even though they are quite small
compared to the diagonal components.

The diagonal components of the weight factor,
$\omega_{nn}$, are shown in the right panel of \Fig{WeightFactor}.
In the figure, both the best copy and the first copy results
are drawn.
We find that the Gribov copy effects can be seen at small $n$
and the two curves approach each other as $n$ increases.
For the lowest component $\omega_{11}$,
the ratio of the first copy result to the best copy one is
about 2.7, that is, the systematic errors due to the Gribov copies
reach 170\%.

Figure \ref{WeightFactor} shows that the lowest component $\omega_{11}$
takes a significantly large value.
For instance, $\omega_{22}$ is about $1/4$ of $\omega_{11}$, and
the diagonal component for $n=20$ is smaller than $\omega_{11}$
by a factor about 100.
It implies that the lowest ghost eigenmode dominates the color-Coulomb potential.
In the Gribov-Zwanziger scenario,
a typical configuration lies near the Gribov horizon and
the lowest FP eigenvalue goes to zero faster than the free field case
as the lattice volume increases.
Since the weight factor $\omega_{nm}$ contains
the inverse of the FP eigenvalue twice,
it is the expected result that
the lowest component $\omega_{11}$ takes a significantly large value
and the lowest FP eigenmode dominates the spectral sum
of the color-Coulomb potential.  
  
\subsection{Correlation of the FP eigenmodes}
\label{sec:Correlation}

\begin{figure}[htbp]
\begin{center}
\resizebox{.46\textwidth}{!}
{\includegraphics{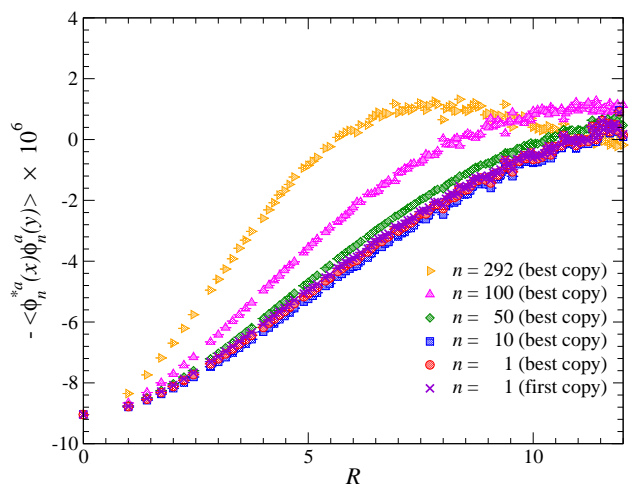}}
\end{center}
\caption{
The correlation function of the FP ghost eigenfunctions,
-$\langle \phi^{\ast a}_n(\vec{x})\phi^a_n(\vec{y}) \rangle$,
is shown as a function of
$R=|\vec{x}-\vec{y}|$.
}
\label{FPcorrelation}
\end{figure}

The correlation function
of the FP ghost eigenmodes,
-$\langle \phi^{\ast a}_n(\vec{x})\phi^a_n(\vec{y}) \rangle$,
is shown in \Fig{FPcorrelation}
as a function of $R=|\vec{x}-\vec{y}|$.
As we have mentioned in \Sec{sec:SpectralSum},
the distance dependence of the color-Coulomb potential
comes from the correlation function.
We see that the correlation function of the low-lying FP eigenfunctions shows
a linearly rising behavior.
The distance dependence of the correlation function
does not so much differ for $n \le 50$.
The confining behavior of the color-Coulomb potential is ascribed
to the fact that the correlation function of the low-lying modes rises linearly
and the corresponding $w_{nm}$ take large values.
Furthermore, we observe that the correlation function for $n=292$
starts to decrease at a large distance; namely,
the correlation function completely loses the confining property
in the sense that it does not rise with distance.
This result strongly supports our expectation that the higher eigenmodes
that are not taken into account in our analysis do not change
the long-range behavior of the color-Coulomb potential and
the color-Coulomb string tension is almost saturated by the low-lying eigenmodes.

In addition to the distance dependence, we notice that the correlation
function is not sensitive to the Gribov copy effects.
This can be seen in \Fig{FPcorrelation} where the first copy and
the best copy results are shown by red circles and violet crosses, respectively.
Therefore, the Gribov copy effects on the color-Coulomb potential
stem from that on the FP eigenvalues and the weight factor.

\section{Summary}
\label{sec:Summary}

In this paper we discussed the effects of the Gribov copies
on the FP eigenvalues and the essential role
of the low-lying FP eigenmodes on the confining color-Coulomb potential
in $SU(3)$ Coulomb gauge Yang-Mills theory
using lattice Monte Carlo simulations.

The low-lying FP eigenvalues are sensitive to the Gribov copies
and the discrepancy between the first copy and the best copy results
exceeds 10\% for the lowest eigenvalue.
The volume scaling of the low-lying eigenvalues was investigated
and we found that the lowest eigenvalue approaches zero much faster than
that in the Landau gauge.

We exploited the fact that the color-Coulomb potential can be expressed as
the spectral sum of the FP ghost eigenmodes.
It was shown that the long-distance linearity of the color-Coulomb potential
is ascribed to the low-lying FP eigenmodes.
Changing $n_{\textrm{max}}$ does not alter the color-Coulomb string tension
for $n_{\textrm{max}} \ge 200$ within the statistical errors, and
about 300 eigenmodes almost saturate the string tension.
The ratio of the color-Coulomb potential to the Wilson string tension
takes values raging from 1.6 to 2.3.
This is consistent with the Zwanziger's inequality.
Moreover, the contribution of the lowest eigenmode to the string tension
is comparable to the Wilson string tension.
Since the inclusion of the higher eigenmodes that have not been taken into account
does not reduce the string tension, our estimated values of the string tension
give a lower limit for the color-Coulomb string tension.
Thus, our result excludes the possibility that the color-Coulomb string tension
saturates the Wilson string tension.

We observed that the weight factor is also sensitive to the Gribov copies.
The systematic errors on the lowest component due to the copies
are shown to be about 170\%.
Although the Gribov copy effects are large,
the lowest component takes a quite large value compared to
the higher components.
This leads to the fact that the lowest eigenmode has a substantial
contribution to the spectral sum of the color-Coulomb potential.

The correlation function of the FP eigenfunctions is shown
to be insensitive to the Gribov copies.
Accordingly, the Gribov copy effects on the color-Coulomb potential
stem from that on the FP eigenvalues.
The correlation function of the low-lying eigenmodes increases with the distance.
By contrast, the correlation function for $n=292$ decreases at large distances.
Therefore, such eigenmodes do not account for the confining property
of the color-Coulomb potential.
This supports our finding that the color-Coulomb string tension
is not altered so much by including the eigenmodes higher than $n \sim 200$.
Our results strongly support the Gribov-Zwanziger scenario:
The infrared dynamics of Yang-Mills theory is governed by the configurations
near the Gribov horizon where the lowest eigenvalue vanishes.

\section{Acknowledgments} 

The simulation was performed on
NEC SX-8R at RCNP,
and NEC SX-9 at CMC, Osaka University.
We appreciate the warm hospitality and support of the RCNP administrators.
Y. N. is supported by a JSPS Grant-in-Aid
from the Ministry of Education, Culture, Sports, Science
and Technology of Japan.
The work is partially supported by
a Grant-in-Aid for Scientific Research by
Monbu-kagakusyo, No. 20340055.

\appendix

\section{Link-link correlator and the Color-Coulomb potential}
\label{app:linklink}

In this Appendix, we present the best copy results for
the color-Coulomb potential in the color-singlet channel
obtained by measuring the correlator of the temporal links.
The necessary equations can be found in
\cite{Nakagawa:2006fk,NakagawaY:PRD77:2008,NakamuraA:PTP115:2006},
and various parameters of the gauge configurations are given
in Table \ref{tab:setup_ccpotential}.

\begin{table}[h]
\caption{
The lattice couplings, the lattice volumes, the number
of configurations used to calculate the color-Coulomb potential.
$N_{cp}$ refers to the number of random gauge copies
generated for each time slice to investigate the Gribov copy effects.
}
\begin{center}
\begin{tabular}{ccccccc}
\hline \hline
$\beta$ & $L^4$ & $a^{-1}$ [GeV] &
$a$ [fm] & $V$[fm$^4$] & \# of confs. & $N_{cp}$ \\
\hline
6.00 & 12$^4$ & 2.118 & 0.0931 & 1.12$^4$ & 100 & 30 \\
6.00 & 16$^4$ & 2.118 & 0.0931 & 1.49$^4$ & 100 & 30 \\
6.00 & 20$^4$ & 2.118 & 0.0931 & 1.86$^4$ & 100 & 30 \\
6.00 & 24$^4$ & 2.118 & 0.0931 & 2.23$^4$ & 100 & 30 \\
\hline \hline
\end{tabular}
\label{tab:setup_ccpotential}
\end{center}
\end{table}

\begin{figure*}[htbp]
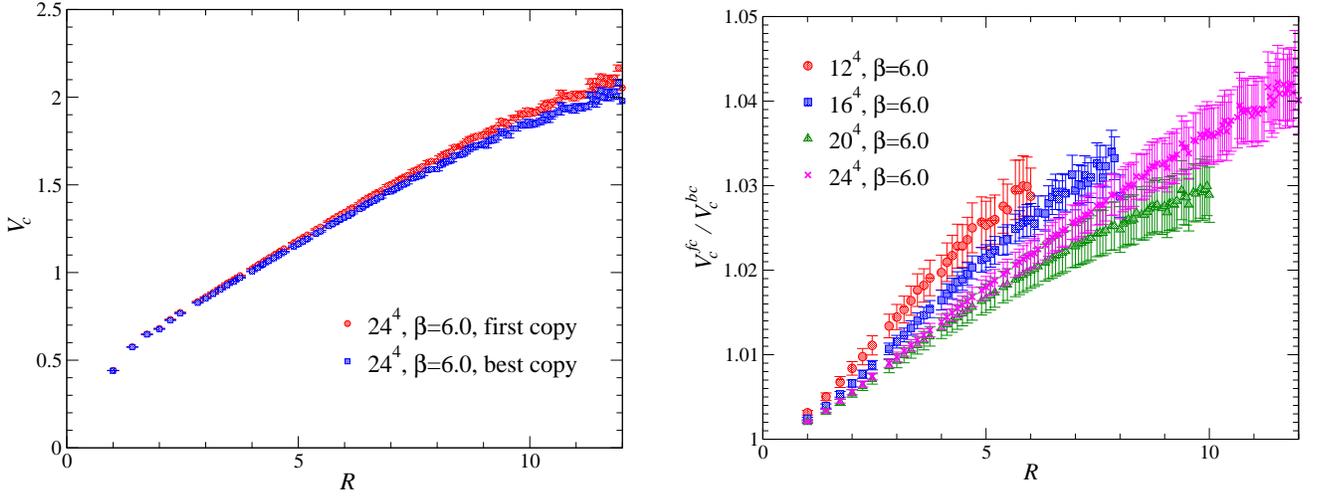

\begin{minipage}{0.46\hsize}\begin{center}
\resizebox{1.\textwidth}{!}
{\includegraphics{fig9a}}
\end{center}\end{minipage}
\hspace{0.03\hsize}
\begin{minipage}{0.46\hsize}\begin{center}
\resizebox{1.\textwidth}{!}
{\includegraphics{fig9b}}
\end{center}\end{minipage}
\caption{
(Left) The color-Coulomb potential in the color-singlet channel
on the $24^4$ lattice at $\beta=6.0$ for the first copies and the best copies.
(Right) The ratio $V_c^{fc}/V_c^{bc}$ of the color-Coulomb potential
on various lattice sizes at $\beta=6.0$.
}
\label{Vcoul_FBratio}
\end{figure*}

The left panel of \Fig{Vcoul_FBratio} shows the color-Coulomb potential
for the best and the first copies on the $24^4$ lattice at $\beta=6.0$,
and the right panel illustrates the ratio $V_c^{fc}/V_c^{bc}$
of the color-Coulomb potential as a function of the distance
on various lattice sizes at $\beta=6.0$.
$V_c^{fc}$ refers to the color-Coulomb potential for the first copies
and $V_c^{bc}$ for the best copies.
We see that the first copies overestimate the color-Coulomb potential.
The deviation is a few percent
while the effects of Gribov copies increase with distance.
The ratio $V_c^{fc}/V_c^{bc}$ decreases
with increasing the lattice volume up to the $20^4$ lattice,
although we do not see such a tendency on the $24^4$ lattice.
It should be noted that the Gribov copy effects are
milder than those on the color-Coulomb potential obtained by
the spectral sum of the FP eigenmodes or by inverting the FP operator.
Further studies are requisite to elucidate why the FP eigenvalues
are tremendously affected by the Gribov copies but the correlator
of the temporal links are not.

We perform a two-parameter fit to extract the string tension
of the color-Coulomb potential,
\begin{equation}
V_c(R) = c + K_c R - \frac{\pi}{12R},
\end{equation}
in the range $4 \le R \le 7$ and find
\begin{equation}
c = 0.502(6), \quad K_c = 0.1429(18), \quad \chi^2/N_{\textrm{DF}} = 0.082.
\end{equation}
The fitting without the L\"uscher term, $-\pi/12R$, gives
\begin{equation}
c = 0.399(5), \quad K_c = 0.1527(17), \quad \chi^2/N_{\textrm{DF}} = 0.146.
\end{equation}
The color-Coulomb string tension is about 3 times
larger than the Wilson string tension,
$K_W = a^2\sigma_W = 0.0513(25)$
\cite{BaliGS:PRD47:1993},
in agreement with the fact that the color-Coulomb potential
provides an upper bound for the static Wilson potential.


\bibliographystyle{h-physrev4}


\end{document}